\documentclass[aps,nofootinbib,preprintnumbers]{revtex4}

\usepackage{amsmath,amssymb}
\usepackage{graphicx,tabularx}

\newcommand{\ie}{{\it i.e.}\;}

\begin{document}

\preprint{MPP-2005-80}

\title{The Quartic Higgs Coupling at Hadron Colliders}

\author{Tilman Plehn\footnote{Heisenberg Fellow}}
\author{Michael Rauch}
\affiliation{Max Planck Institute for Physics, Munich, Germany}

\begin{abstract}
  The quartic Higgs self-coupling is the final measurement in the Higgs
  potential needed to fully understand electroweak symmetry breaking. None of
  the present or future colliders are known to be able to determine this
  parameter.  We study the chances of measuring the quartic self-coupling at
  hadron colliders in general and at the VLHC in particular.  We find the
  prospects challenging.
\end{abstract}
  
\maketitle


The LHC and a future linear collider are widely regarded as an ideal
combination of experiments to understand electroweak symmetry breaking, \ie
study the Higgs boson and measure its couplings to all Standard Model bosons
and fermions.  According to the electroweak precision data~\cite{lepewwg} we
expect to discover and identify a light Standard Model Higgs boson at the
LHC~\cite{discovery,vertex,duehrssen}. At the ILC we will be able to measure
Higgs couplings to all Standard Model particles with great
precision~\cite{ilc}.~\footnote{Recently, it has also been shown that in
  supersymmetric models the Higgs couplings to neutralinos and charginos can
  be determined at linear colliders to high precision~\cite{split_susy}.} A
particularly exciting task is the measurement of the trilinear Higgs
self-coupling: for a Standard Model Higgs boson heavier than 150~GeV this
coupling can be measured at a luminosity-upgraded LHC (but not at the
ILC)~\cite{self_psz,self_basics,self_lhc}.  In contrast, for small Higgs
masses around 120~GeV it can be measured at the ILC (but possibly not at the
LHC)~\cite{maggi, self_ilc, self_rare}.\bigskip

\underline{Higgs potential at Colliders:} The measurement of the Higgs
self-coupling is crucial to determine the Higgs potential --- the way
we think the electroweak symmetry is broken.  A general
parameterization of the Higgs potential with one doublet $\Phi$ (as in the
Standard Model) is~\cite{higgs_pot}:
\begin{equation}
\label{eq:pot}
V(\Phi) \, = \,
             \sum_{n \geq 0} \frac{\lambda_n}{\Lambda^{2n}}
             \left( \Phi^\dagger\Phi - \frac{v^2}{2} 
             \right)^{2+n}
        \, = \,
             \lambda_0
             \left( \Phi^\dagger\Phi - \frac{v^2}{2} 
             \right)^2 + {\cal O}\left( \frac{1}{\Lambda^2} \right),
\end{equation}
where $v=(\sqrt{2}G_F)^{-1/2}$ is the vacuum expectation
value, and $G_F$ is the Fermi constant. In the Standard Model,
$\lambda_0=\lambda_{SM}=m_H^2/(2v^2)$. If we consider the Standard Model an
effective theory, $\lambda_0$ stands for two otherwise free parameters, namely
the trilinear and the quartic scalar self couplings. An upper limit can be
determined using unitarity arguments, assuming the model's validity to high
energy scales~\cite{unit}. In the Standard Model, the two self-couplings are
linked as the leading terms in eq.(\ref{eq:pot}), namely $\lambda_3/\lambda_4
= v$, and higher dimension operators are not expected to appear much below the
Planck scale. If we allow for an intermediate scale $\Lambda \ll M_{\rm
  Planck}$ and include the higher dimensional terms $n=1,2$ both self
couplings receive different corrections: $\lambda_3 \to \lambda_3 [1 +
\lambda_1 v^2/(\lambda_0 \Lambda^2)]$ and $\lambda_4 \to \lambda_4 [1 + 6
\lambda_1 v^2/(\lambda_0 \Lambda^2) + 4 \lambda_2 v^4/(\lambda_0 \Lambda^4)]$.
In general, it is not even guaranteed that both self-couplings have to be
positive, since the stability of the general Higgs potential is guaranteed by
the sign of the highest power in the Higgs field alone.

\begin{figure*}[b] 
\includegraphics[width=4.4cm]{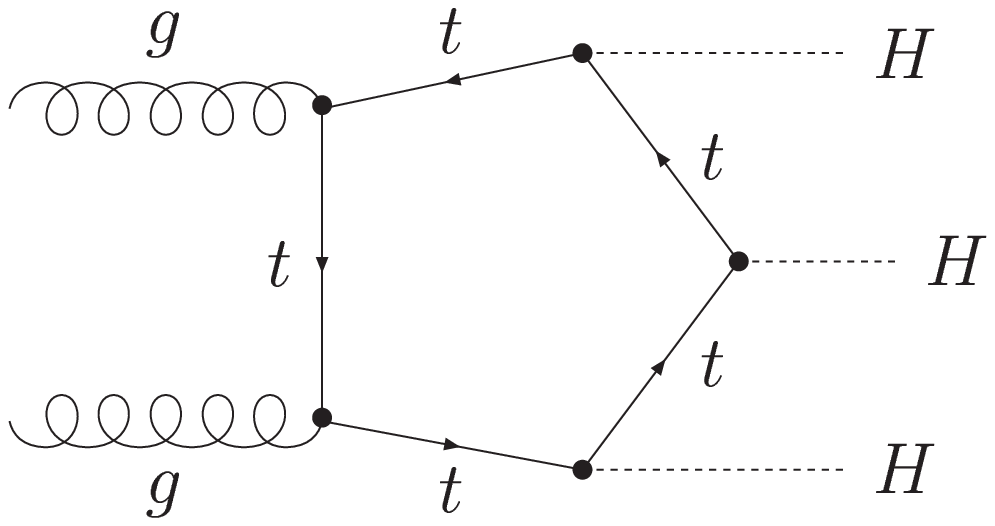} 
\includegraphics[width=4.4cm]{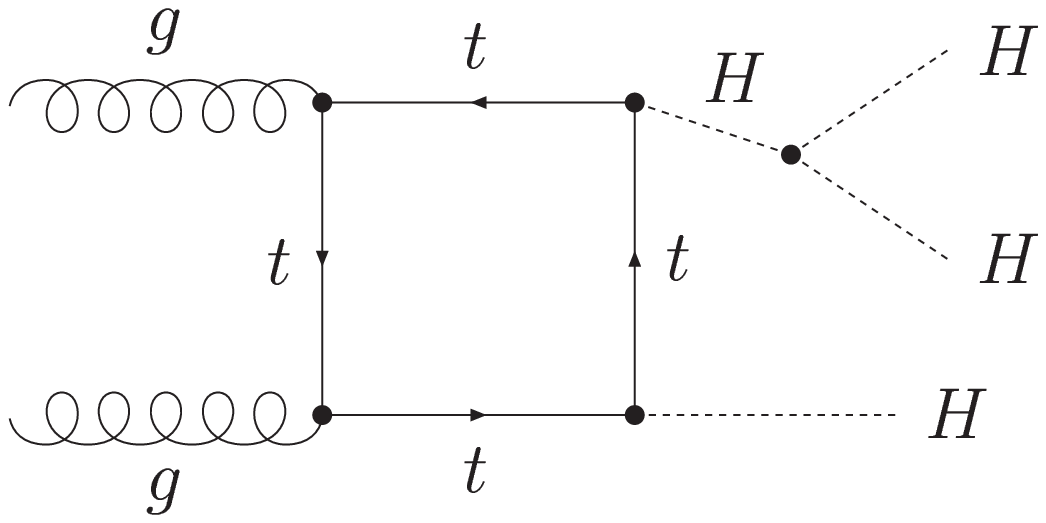} 
\includegraphics[width=4.4cm]{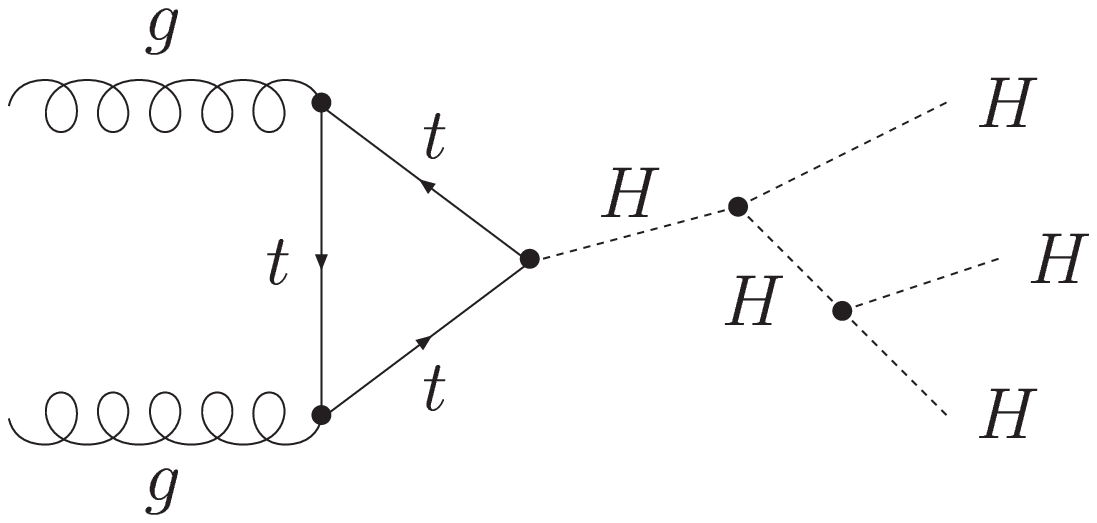} 
\includegraphics[width=4.4cm]{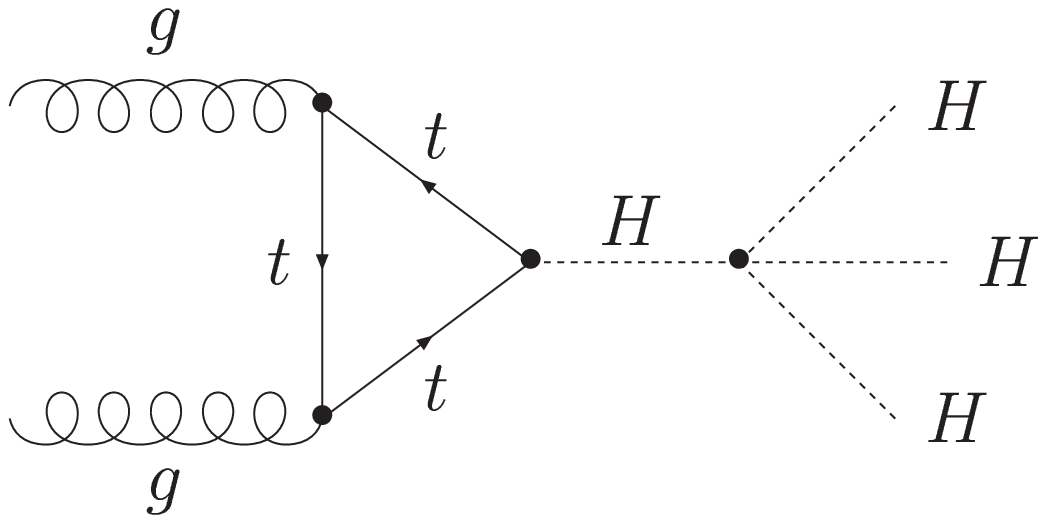} 
\caption[]{\label{fig:feyn} 
  Examples of Feynman diagrams contributing to the process $gg \to HHH$.}
\end{figure*}

The relation between the Higgs mass and each self-coupling as well as between
the two self-couplings can change dramatically when we move to the MSSM with 
its two Higgs doublets.  If we replace the Standard Model Higgs with the
light CP-even scalar $h^0$ the relation between the self coupling becomes
$\lambda_{3h}/\lambda_{4h} = v \; \sin(\beta+\alpha)/\cos 2
\alpha$~\cite{mssm}. As usual, $\tan \beta$ is the ratio of the two vacuum
expectation values and $\alpha$ is the mixing angle between the two Higgs
scalars.  However, if we assume a mass hierarchy between the light Higgs
scalar and the remaining Higgs sector the difference to a Standard-Model like
Higgs is very small.\smallskip

The measurement of the trilinear Higgs coupling requires the production of at
least two Higgs bosons. The gluon fusion process at the LHC can proceed
through a box-shaped top loop or through a triangular top loop and an
intermediate Higgs boson. While it is well known that the total rate can be
well approximated by the heavy-top
approximation~\cite{self_psz,self_basics} (for three Higgs production see Ref.~\cite{vdbij}) it has been shown for Higgs pair production that the
distributions in particular in the threshold region will come out completely
wrong~\cite{self_lhc}. Precisely this threshold behavior carries the information on the Higgs
self-coupling.\smallskip

In this brief letter we study a possible measurement of the quartic
Higgs self-coupling at hadron colliders. In analogy to the measurement
of the trilinear coupling we now produce three Higgs bosons in gluon
fusion. The four topologies shown in Fig.~\ref{fig:feyn} will appear: (a) continuum production of three
Higgs bosons through a pentagon top-loop, (b) the production of two
Higgs bosons with a subsequent decay via the trilinear self-coupling,
and finally the production of one Higgs boson with a decay through
either (c) two three-Higgs vertices or (d) through one quartic
self-coupling.  From this list it is obvious that it will not be
possible to make any statement about the quartic self-coupling without
having information on the trilinear coupling and the top Yukawa coupling~\cite{tth_lhc}. Moreover, it is fairly
obvious that the LHC even including a major luminosity upgrade will
not be able to supply enough three-Higgs events. Instead, we ask the
question what a future 200~TeV VLHC~\cite{vlhc} could do, keeping in
mind that even future high-energy linear colliders like CLIC will not
be able to measure this coupling~\cite{maggi,clic}. In other words, we
are trying to determine if {\sl any} future high-energy collider will
be able to completely measure the parameters of the Higgs
potential~(\ref{eq:pot}).\bigskip

\begin{figure*}[t] 
\includegraphics[width=6.0cm,angle=270]{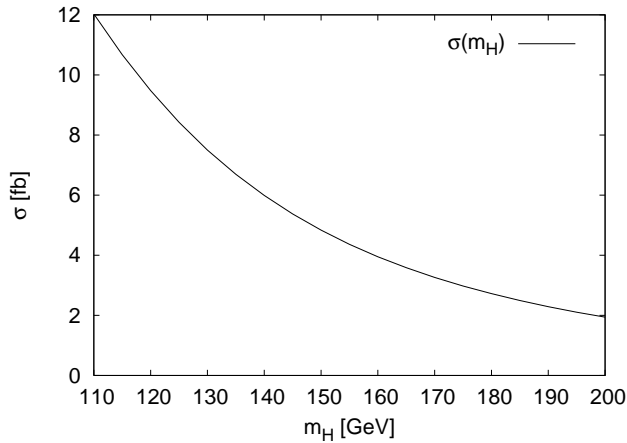}
\caption[]{\label{fig:tot} 
  Total cross section for the production process $gg \to HHH$ at the
  200~TeV VLHC in the Standard Model.}
\end{figure*}

\begin{figure*}[t] 
\includegraphics[width=6.0cm,angle=270]{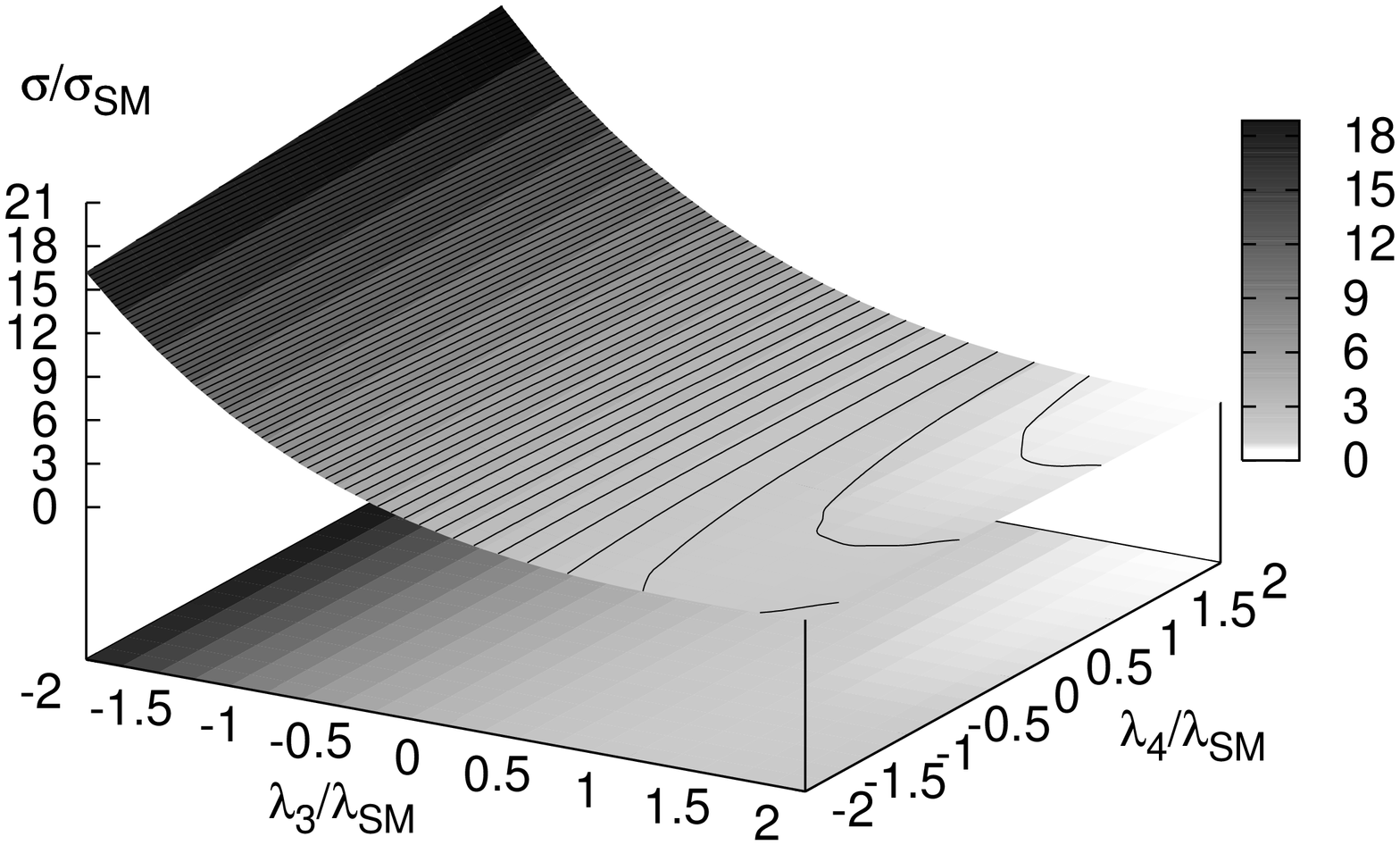}
\hspace*{3mm}
\includegraphics[width=6.0cm,angle=270]{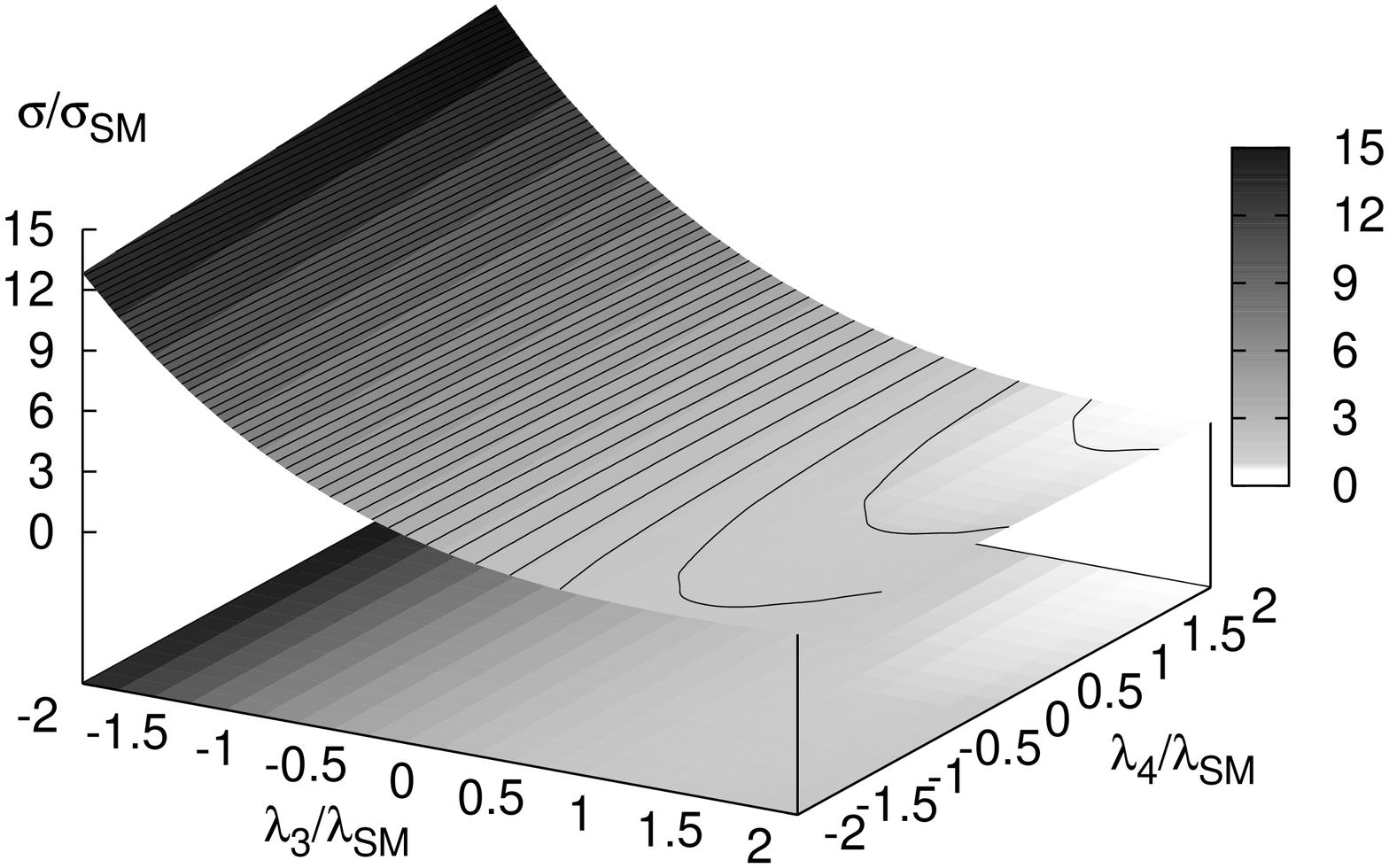}\\
\includegraphics[width=5.5cm,angle=270]{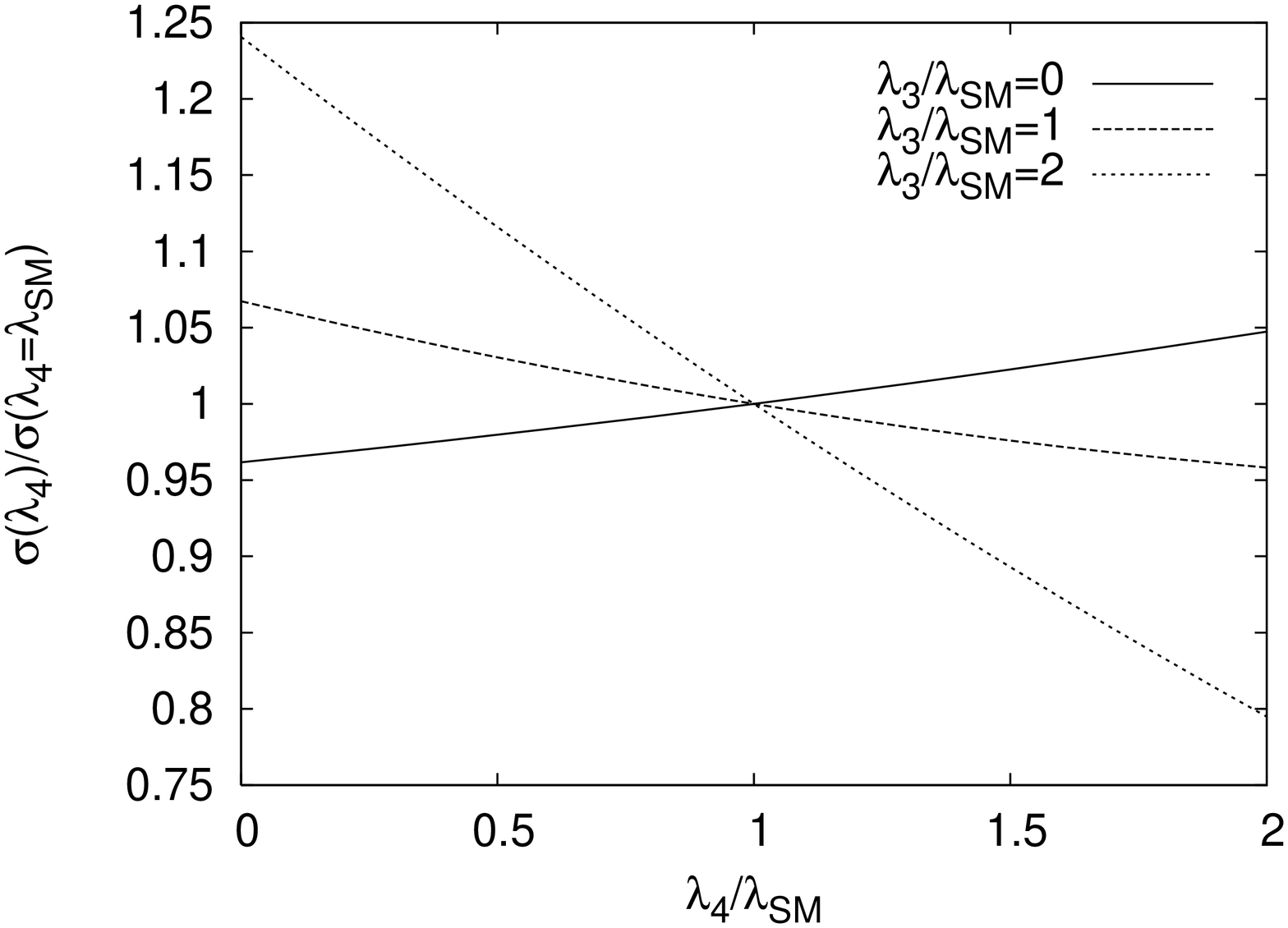}
\hspace*{6mm}
\includegraphics[width=5.5cm,angle=270]{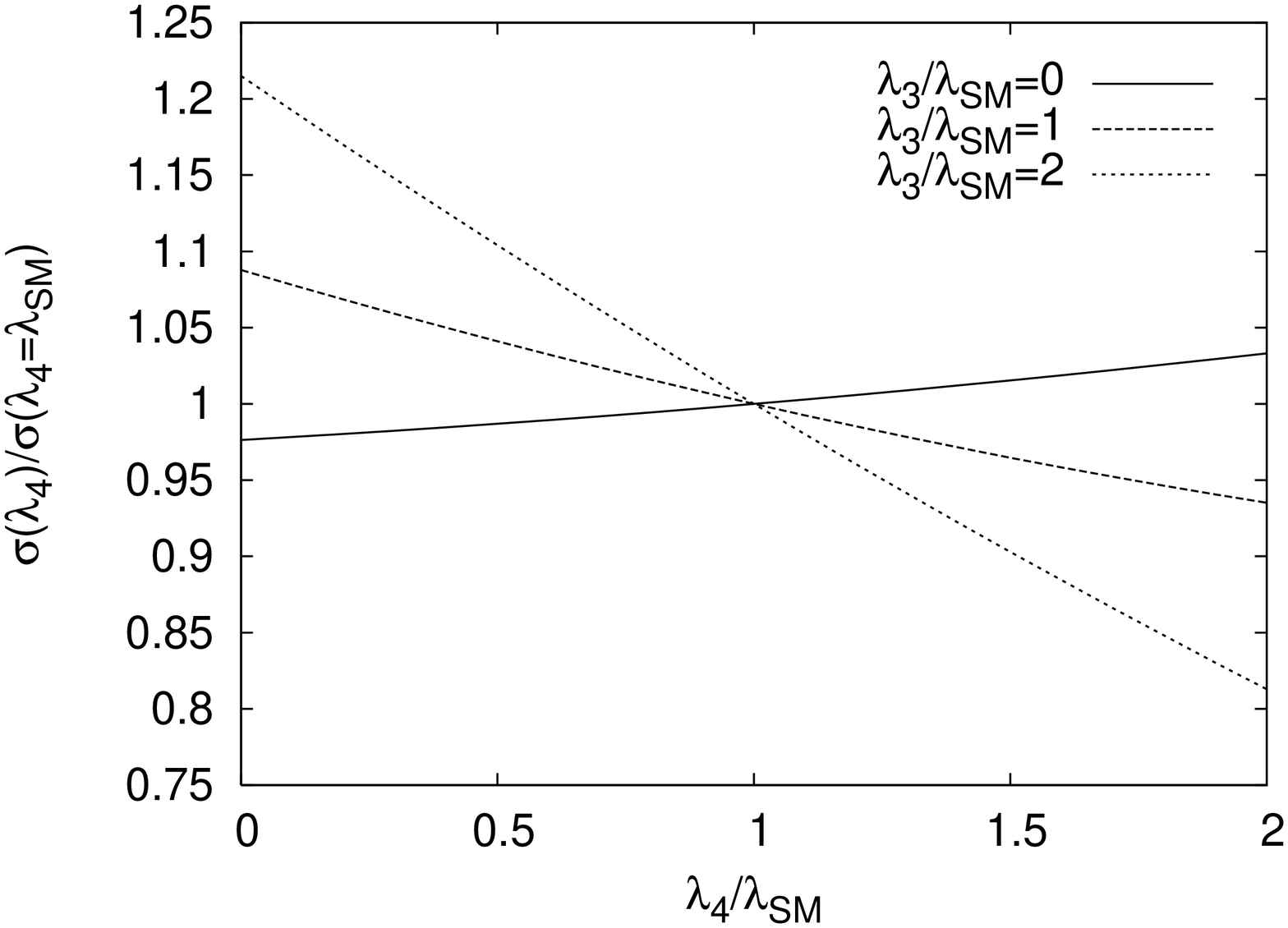}
\caption[]{\label{fig:3d} 
  Total cross section ratios normalized to the Standard Model values
  for the production process $gg \to HHH$ at the LHC~(left) and at the
  200~TeV VLHC~(right). The Higgs mass is fixed to 120~GeV, the
  absolute values of triple and quartic self-couplings are varied up
  to twice the Standard Model values. Contour lines are included every
  0.25 steps on the z axis. In the lower set of figures the triple
  Higgs self-coupling is fixed. The Standard Model values for the
  cross sections are $6.25\cdot10^{-2}$~fb at the LHC and 9.45~fb 
  at the VLHC.}
\end{figure*}

\underline{Three--Higgs production:} The one-loop diagrams
contributing to the process $gg \to HHH$ are leading order, \ie they
are finite for any value of the Higgs self-couplings. We compute the
total and differential cross sections using the HadCalc
program~\cite{hadcalc}. The Feynman diagrams are constructed using
FeynArts~\cite{feynarts}, the matrix elements are calculated by
FormCalc~\cite{formcalc}, and the loop integrals are numerically
evaluated using LoopTools~\cite{looptools}, where we have added the
scalar five-point function~\cite{fivepoint} and modified the general
four-point function~\cite{niere}. For the top mass we use the
on-shell value ($m_t$ = 178~GeV), because it has been shown to lead to
perturbatively stable cross section predictions for the single-Higgs
production through a one-loop amplitude~\cite{spirix}.  The bottom
loops are included in our numerical analysis, but their effect is
below one per cent.

We show the total cross section for the production of three Standard
Model Higgs bosons at a 200~TeV VLHC in Fig.\ref{fig:tot}. The cross
sections are quoted without branching fractions, acceptance cuts, or
efficiencies. 
In Fig.~\ref{fig:3d} we show the dependence of the LHC and the 200~TeV
VLHC cross sections on the trilinear ($\lambda_3$) and quartic
($\lambda_4$) Higgs self couplings.  The central values at the LHC and at the VLHC are
$6.25\cdot10^{-2}$~fb and 9.45~fb (Standard Model couplings). The
fact that $\lambda_3$ contributes to many more topologies than
$\lambda_4$ (with its single diagram) is reflected in the much steeper
behavior of the total cross sections as a function of $\lambda_3$ than
as a function of $\lambda_4$: each of the three topologies (triangle,
box, pentagon) alone would yield a rate of $(0.46, 8.20, 17.07) \cdot
10^{-2}$~fb at the LHC. If we compute only the propagator-suppressed
triangle contribution and keep either $\lambda_3 \neq 0$ or $\lambda_4
\neq 0$, we are left with $0.17\cdot 10^{-2}$~fb from the trilinear
self-coupling and with $0.08\cdot 10^{-2}$~fb from the quartic
self-coupling, with a constructive interference. This means that the
contribution from the quartic self-coupling is suppressed by almost
two orders of magnitude.\medskip

The interference between the continuum and the box is indeed
destructive (as we would expect from Ref.~\cite{self_lhc}), which is
the primary reason for the steep decline with $\lambda_3$ shown in
Fig.~\ref{fig:3d}. The interference between the continuum and the
triangle diagrams is constructive, but because of the more similar
kinematic configuration the destructive interference between the box
diagrams and the triangle diagrams leads to the slight decrease of the
cross section with growing $\lambda_4$. If we switch off the box
contributions ($\lambda_3=0$) the constructive interference between
continuum and triangle topologies switches around the behavior of the
total cross section as a function of $\lambda_4$. This behavior can be
understood analytically in the limit $m_t \gg m_H$, using the
low--energy theorem for the leading form factors in $m_H/m_t \;
(\hat{s} \sim m_H^2)$~\cite{low_energy}\footnote{We use the conventions
  as in Ref.~\cite{self_psz}. The form factor is basically the matrix
  element squared without couplings or additional $s$-channel
  propagators. The top Yukawa coupling and the top mass in the
  propagators are both denoted as $m_t$.}. These leading form factors
for an increasing number of external Higgs scalars can be iteratively
derived from the top loop in the gluon self-energy: $F_{(n+1)H} =
m^2_t \partial (F_{nH}/m_t)/\partial m_t$.  We obtain $F_{\rm
  triangle} = -F_{\rm box} = F_{\rm pentagon} = 2/3 + {\cal
  O}(m_H^2/m_t^2)$. This relative sign explains the structure of the
constructive and destructive interferences observed above.\smallskip

Because of the fairly strong dependence on $\lambda_3$, where the cross
section decreases for increasing $\lambda_3$, we will be able to exclude the
hypothesis of $\lambda_3 = 0$ most easily, exactly the same way as in the case
of Higgs pair production~\cite{self_lhc}. The same trick could work for the
quartic self-coupling $\lambda_4$ as long as $\lambda_3 \gtrsim \lambda_3^{\rm
  SM}$.  However, the typical variations in the total rate are above 100~\%
for varying $\lambda_3$ and less than 20~\% for varying $\lambda_4$. This
means that after including the systematic errors on the cross section (like
higher-order QCD uncertainties) and the statistical error on the measurement
of $\lambda_3$ there is little hope to see an effect of $\lambda_4$ in the
total rate at the LHC or at the VLHC.\medskip

\begin{figure*}[t] 
\includegraphics[width=6.0cm,angle=270]{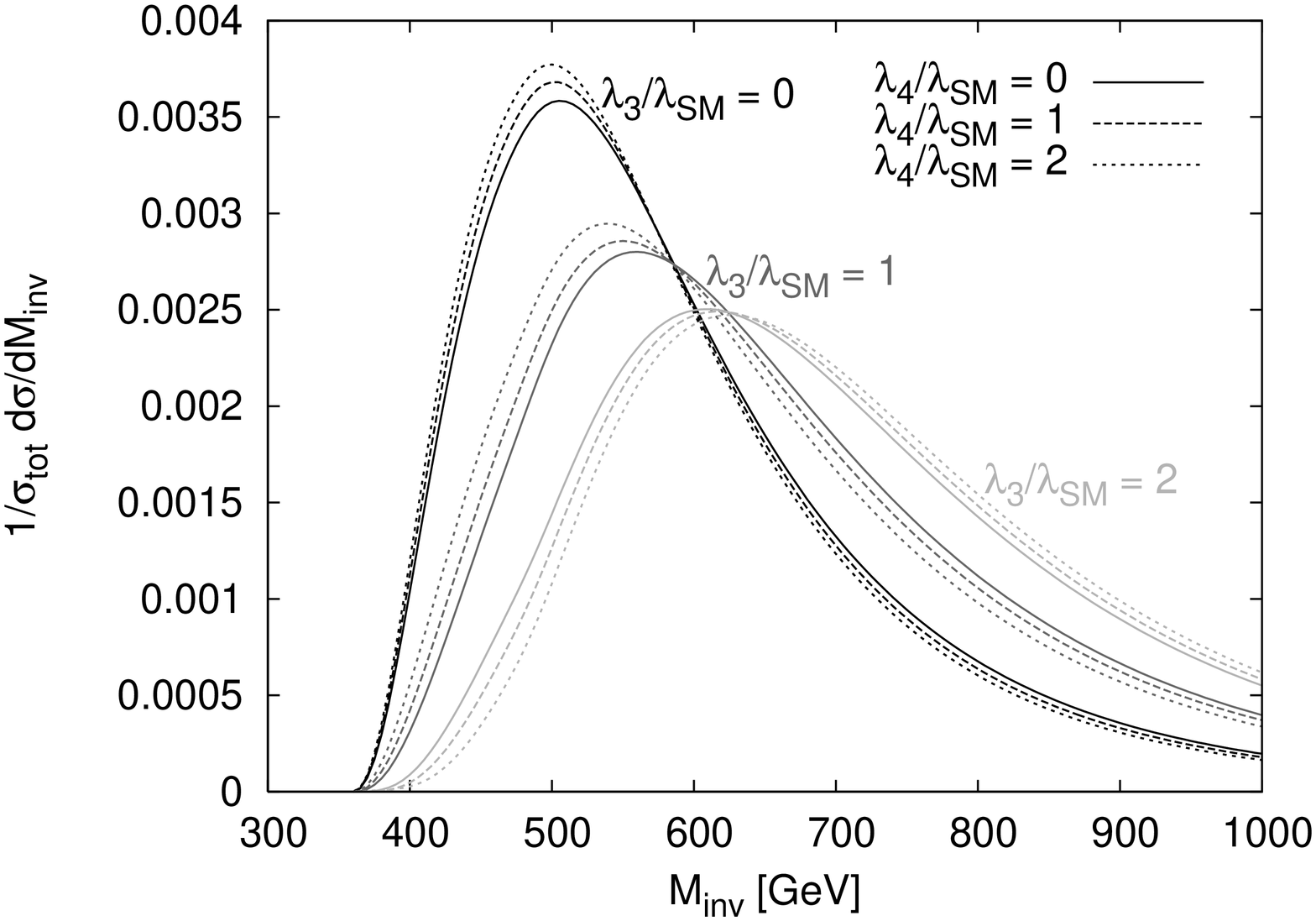} 
\hspace*{3mm}
\includegraphics[width=6.0cm,angle=270]{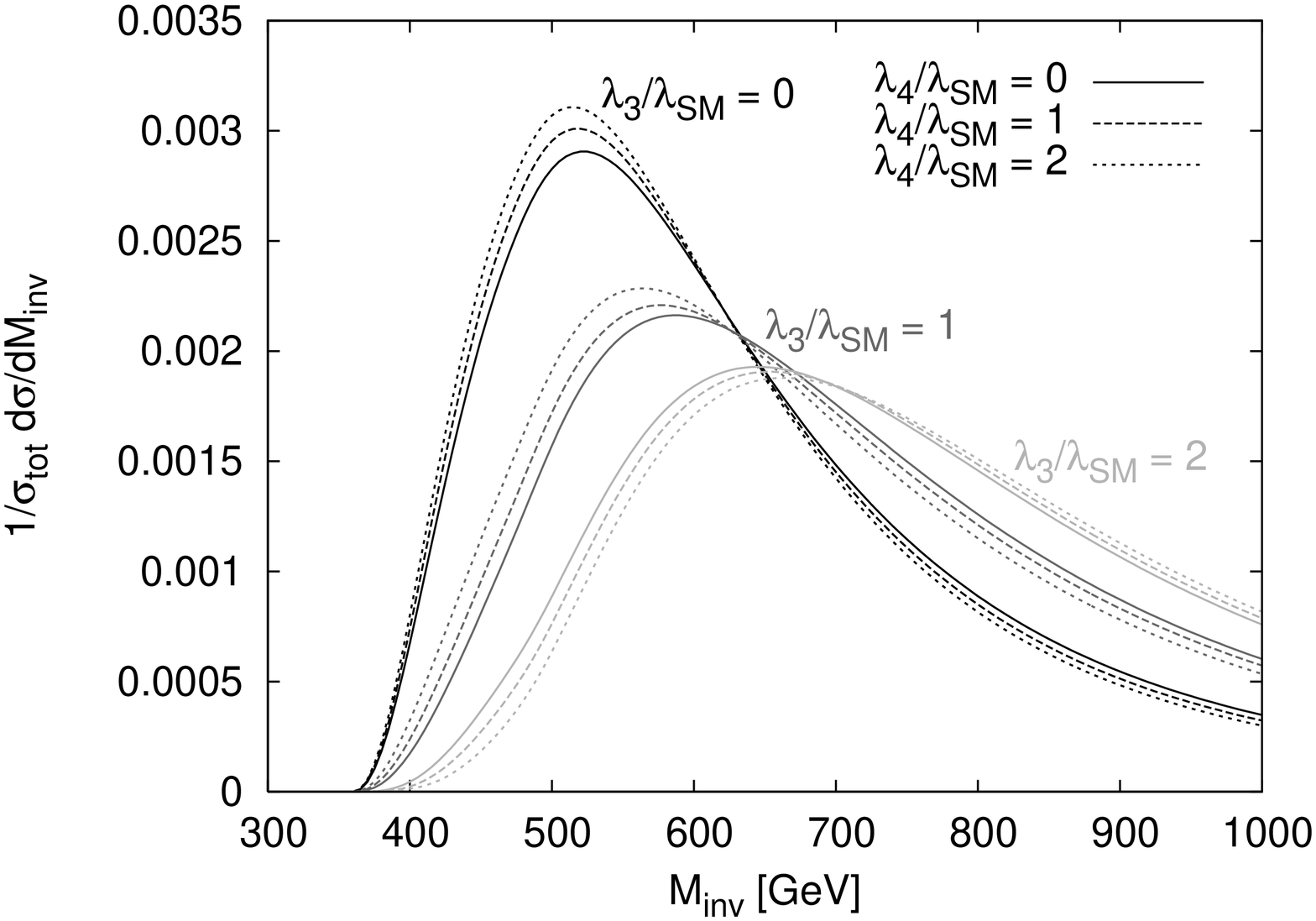}
\caption[]{\label{fig:distri} 
  Normalized partonic center-of-mass energy distributions for $gg \to
  HHH$ at the LHC~(left) and at the 200~TeV VLHC~(right). The Higgs
  mass is fixed to 120~GeV. The trilinear and quartic Higgs
  self-couplings are varied independently.}
\end{figure*}

Remembering what we know about the measurement of the trilinear self-coupling
at the LHC we can compute the normalized distribution of the partonic
center-of-mass energy. At the LHC its shape will have maximum discriminating
power, because of the interference effects between continuum and intermediate
Higgs topologies at threshold~\cite{self_lhc}. We show these distributions in
Fig.~\ref{fig:distri}, again varying $\lambda_3$ and $\lambda_4$
independently. Indeed, we see that there is a sizeable shift when we vary
$\lambda_3$ while the numerical impact of $\lambda_4$ is about an order of
magnitude smaller. We also see that for $\lambda_3=0$ the order of the three
curves for $\lambda_4/\lambda_{\rm SM}=0,1,2$ turns around together with the
sign of the interference. While this distribution will be the key to the LHC
measurement of $\lambda_3$, the curves in Fig.~\ref{fig:distri} look
challenging for a measurement of $\lambda_4$, in particular once we again
include the error on the measurement of $\lambda_3$.\bigskip

\underline{Summary:} To measure the entire set of parameters in the
Higgs potential (and completely understand electroweak symmetry
breaking) we have to measure the quartic Higgs self-coupling. At the
LHC and the ILC the measurement of the trilinear self-coupling will
already be a task which requires large luminosities and a very good
understanding of the
detector~\cite{self_lhc,self_rare,maggi,self_ilc}. We know that CLIC
would not be able to measure the quartic Higgs coupling~\cite{clic}, so
what is left is the high-energy mode of the VLHC. To get a rough idea
if the quartic coupling might be visible we compute the total cross
section as well as the partonic center-of-mass energy distribution for
the process $gg \to HHH$.  For a 120~GeV Higgs the cross section at
the VLHC is 9.45~fb, so we might even be able to observe triple Higgs
production. However, based on this simple study without decays or
detector effects we conclude that the measurement of the quartic Higgs
self-coupling will be seriously challenging due to the interference
structures between the different topologies contributing to the
process. Moreover, a measurement of the quartic self-coupling requires
a very good knowledge of the value of the trilinear self-coupling,
which at the moment is not (yet) established.\bigskip
 
\underline{Acknowledgments:} We would like to thank W.~Hollik for his 
support and careful reading of the manuscript, W.~Kilian
for helpful discussions and S.~Dittmaier for crucial technical help
and discussions concerning the loop integrals. 
Moreover, we would like to thank U.~Baur and D.~Rainwater for 
very useful comments after reading our manuscript. Last but not least we
are grateful to D.~Zeppenfeld for discussions in an early stage of this
project.


\bibliographystyle{plain}

\end{document}